\documentclass[pra,twocolumn,superscriptaddress,showpacs]{revtex4}

\usepackage{amsbsy}
\usepackage{amsfonts}
\usepackage{amssymb}
\usepackage{amsmath}
\usepackage{color}
\usepackage{graphicx}

\begin{document}

%opening

\title{Cooperativity of a few quantum emitters in a single-mode cavity}

\author{Eduardo Mascarenhas}
\affiliation{Departamento de F\'isica, Universidade Federal de Minas Gerais, C.P. 702, 30123-970, Belo Horizonte, MG, Brazil}

\author{Dario Gerace}
\affiliation{Dipartimento di Fisica, Universit\`a di Pavia, via Bassi 6, I-27100 Pavia, Italy}

\author{Marcelo Fran\c{c}a Santos}
\affiliation{Departamento de F\'isica, Universidade Federal de Minas Gerais, C.P. 702, 30123-970, Belo Horizonte, MG, Brazil}

\author{Alexia Auff\`eves}
\affiliation{Institut N\'eel-CNRS, 38042 Grenoble Cedex 9, France}

\begin{abstract}

We theoretically investigate the emission properties of a single-mode cavity coupled to a mesoscopic number of incoherently pumped quantum emitters. We propose an operational measure for the medium cooperativity, valid both in the bad and in the good cavity regimes. We show that the opposite regimes of subradiance and superradiance correspond to negative  and positive cooperativity, respectively. The lasing regime is shown to be characterized by nonnegative cooperativity. In the bad cavity regime we show that the cooperativity defines the transitions from subradiance to superradiance. In the good cavity regime it helps to define the lasing threshold, also providing distinguishable signatures indicating the lasing regime. Increasing the quality of the cavity mode induces a crossover from the solely superradiant to the lasing regime. Furthermore, we verify that lasing is manifested in a wide range of positive cooperative behavior, showing that stimulated emission and superradiance can coexist. The robustness of the cooperativity is studied with respect to experimental imperfections, such as inhomogeneous broadening and pure dephasing.

\end{abstract}

%\pacs{ 03.65.Yz; 03.67.Hk; 03.67.Pp}

\maketitle

\section{Introduction}
The optical properties of $N$ emitters interacting with the same electromagnetic environment are drastically different from those of $N$ independent emitters, each interacting with its own reservoir. Signatures of cooperative behavior in the spontaneous emission of an atomic ensemble were first discussed in the context of the celebrated superradiance decay \cite{Dicke}, where constructive interference of the atomic dipoles leads to enhanced relaxation of the atomic population, and to the emission of an intense and delayed pulse of light \cite{gross_haroche}. 

More recently, the steady state properties of the light field emitted by such a medium have been theoretically investigated, assuming continuous incoherent pumping, both for what concerns its spectral \cite{poddubny2010,laussy2011prb} and statistical characteristics \cite{Woggon,Holland,Auffeves2011}.  Subradiant and superradiant regimes have been defined and identified, respectively corresponding to the emission of less/more light than in the independent emitters case. In these works, a common electromagnetic environment for the quantum emitters is provided by a leaky cavity, which acts as a collective decay channel. Naturally, increasing the quality factor of the resonator can eventually induce lasing, such that both phenomena could be observed in the same system at different pumping rates, in principle. 

In the weak coupling regime, the concept of cooperativity has been implicitly used in the literature by adiabatically eliminating the cavity mode in the theoretical description, as described, e.g., in Ref.~\onlinecite{Holland}. However, even in this particular case there was no precise definition to measure such a quantity. 
In this paper, we propose an operational measure to quantify the cooperativity of a few quantum emitters coupled to the same single-mode resonator. We show that it works well irrespective of the operating regime of the system (weak or strong coupling, good or bad cavity, etc.). In that sense, it is also universal. We also address the spectral and the statistical properties of the emitted radiation supporting our analysis with the cooperativity measure. 

Specifically, we analyze an ensemble of two-level emitters and study their cooperativity with respect to experimentally addressable parameters. 
We show that the cooperativity assumes negative values at subradiance and positive values at superradiance, which clearly indicates the change of regime. 
As the cavity quality factor is increased, we observe a crossover to a lasing regime, where the cooperativity shows distinctive signatures below and above threshold that allow to identify the nonlasing-lasing crossover.
In the good cavity regime, our measure can reasonably be used to define the lasing threshold (usually not well defined for lasers of few emitters) as the pump rate increases, and the cooperativity changes from negative to non-negative values. Furthermore we show that lasing is manifested in a wide range of positive cooperativity, showing that lasing and superradiance are distinct phenomena that can coexist: when each emitter is strongly coupled to the resonator mode, a \textit{cooperative lasing} regime manifests by a delayed quenching of the laser at strong incoherent pumping. Finally, we analyze the robustness of cooperativity to experimental imperfections, such as inhomogeneous broadening and dephasing.
From an experimental point of view, systems that might realize the present model include small assemblies of artificial atoms, such as semiconductor quantum dots in microcavities \cite{strauf06,kapon2011}, superconducting qubits coupled to microstrip line resonators~\cite{dicarlo2010nat}, or defects in solid state cavities~\cite{faraon2011}.

The paper is organized as follows. In Sec.~\ref{sec:model} we define the hamiltonian and master equation we have used, and discuss the numerical techniques used to solve the problem. In Sec.~\ref{sec:cooperativity} we introduce a novel definition of cooperative fraction, which is able to discriminate between the various regimes of the model. In Sec.~\ref{sec:crossover} we investigate the signatures of the superradiance-lasing crossover as a function of the incoherent pumping rate, while its robustness to external parameters is discussed in Sec.~\ref{sec:parameters}, with an emphasis on the role of the cooperativity as a valuable measurable quantity.

\section{System, model and methods}\label{sec:model}

\begin{figure}[t]
\includegraphics[width=0.75\linewidth]{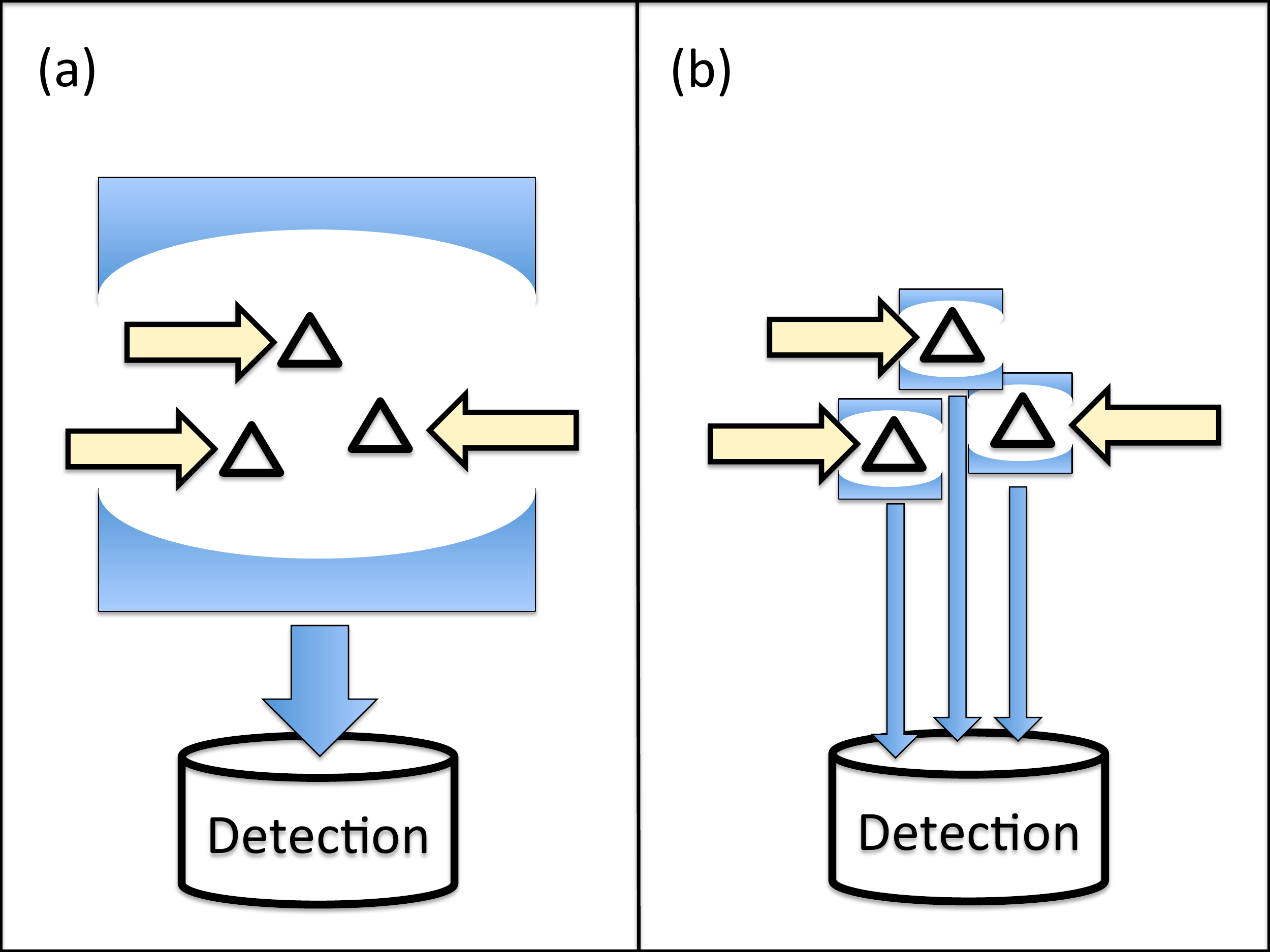}
\caption{The system under investigation, and the operational definition of cooperativity. N emitters being incoherently pumped, two situations are compared. (a) A detector captures the output of a cavity containing the ensemble of emitters. (b) Each emitter is coupled to an individual cavity, and each output signal is assumed to be detected and summed.}\label{CoopsFig}
\end{figure}

The system under study is a single-mode electromagnetic cavity coupled to $N$ two level emitters, as schematically pictured in Fig.~\ref{CoopsFig}a. The emitters are incoherently pumped and may be dephased or detuned from the cavity frequency. The model describing this system is the well known Tavis-Cummings Hamiltonian \cite{TavisCummings}, which is written in a frame rotating at the cavity frequency as
\begin{equation}
H=\sum_i^N[ \delta_i \sigma^{\dagger}_i\sigma_i+g(\sigma_i^{\dagger} a+\sigma_i a^{\dagger})] \, ,
\end{equation}
with $\sigma_i$ being the lowering operator for the $i$th emitter, $a$ the annihilation operator for the cavity mode, $\delta_i$ the detuning of the $i$th atom from the cavity resonance, and $g$ the light-matter coupling constant depending on the two-level system oscillator strength \cite{andreani99prb}. 
The incoherent processes in the system, namely cavity losses, incoherent pumping of the emitters, and their pure dephasing are described within a Markovian approximation and Lindblad dynamics, with the Liouville-von Neumann equation of motion being written as (see, e.g.,~\cite{Carmichael_book} for a rigorous derivation)
\begin{equation}
\dot{\rho}=\mathcal{L}(\rho)=-i[H,\rho]+k \mathcal{D}_a(\rho)+\sum_i[P\mathcal{D}_{\sigma^{\dagger}_i}(\rho)+\gamma\mathcal{D}_{z_i}(\rho)] \, ,
\end{equation}
with $k$ being the rate of photon leakage from the cavity, $P$ the incoherent pumping rate, $\gamma$ the pure dephasing rate, and $z_i=\sigma_i^{\dagger}\sigma_i$. The Lindblad expression for an arbitrary operator, $x$, is given by
\begin{equation}
\mathcal{D}_x(\rho)=-\frac{1}{2}[x^{\dagger}x\rho+\rho x^{\dagger}x]+x\rho x^{\dagger} \, .
\end{equation}

In the following we focus on the stationary properties of the system and numerically compute the steady state values of the cavity population and atomic inversion, respectively. 
In addition, we calculate the second-order coherence function at zero time delay of the cavity field, defined as~\cite{Carmichael_book}
\begin{equation} 
g^2(0)=\frac{\left\langle a^{\dagger}a^{\dagger} aa\right\rangle}{\left\langle a^{\dagger}a\right\rangle^2} \, ,
\end{equation}
and the cavity emission spectrum~\cite{Carmichael_book}
\begin{equation} 
S(\omega)=\int \lim_{\tau\rightarrow\infty}\left\langle a^{\dagger}(t+\tau)a(\tau)\right\rangle e^{i\omega t }dt \, ,
\end{equation}
which is the Fourier transform of the first-order correlation function. The latter can be calculated for the stationary state as
$\left\langle a^{\dagger}(t)a(0)\right\rangle=\mathrm{tr}\left\{ a^{\dagger}e^{\mathcal{L}t}a\rho_s \right\}$,
with $\rho_s$ being the steady state density matrix of the system. 

The determination of the asymptotic state can be numerically done by standard sparse matrix diagonalization of the total Lindblad operator, such that $\mathcal{L}\rho_s=0\rho_s$. In this work, we used a shift-and-invert Arnoldi method, which is coded in the ARPACK library that is built-in the MATLAB environment~\cite{Lmatrix}. The time evolution needed to compute the correlation function is done with the Arnoldi method, which is the optimization of the matrix exponential in the Krylov subspace~\cite{Lmatrix2}. A brief description of the method is as follows. One needs to compute propagations of the form $\rho(t+\Delta t)=e^{\mathcal{L}\Delta t}\rho(t)$, with the major difficulty being the calculation of the matrix exponential. This is done in the Krylov subspace, which is the subspace generated by the iterative application of the Lindblad matrix
\begin{equation}
\mathcal{K}_m[\mathcal{L},\rho(t)]=\mathrm{span}\left\{ \rho(t), \mathcal{L}\rho(t),\mathcal{L}^2\rho(t),...,\mathcal{L}^{m-1}\rho(t)\right\} \, .
\end{equation}
This basis for the Krylov space can be ortho-normalized. In fact, the subspace can be constructed in orthonormal form by the modified Gram-Schmidt method. We call $V$ the matrix whose columns are the orthonormal basis-vectors of the Krylov space. For the calculations shown in the present work, we have always kept 20 Krylov vectors. Thus, the huge numerical problem can be reduced to a small space by projecting the Lindblad into an upper Hessenberg form, $V^{\dagger}\mathcal{L}V=\mathcal{H}$. The eigenvalues of the Hessemberg matrix are Ritz approximate eigenvalues of $\mathcal{L}$, such that $\mathcal{H}=UDU^{-1}$. Finally, we have the solution (explicitly written in the most effective order of multiplication)
\begin{equation}
\rho(t+\Delta t)=V\left[Ue^{D\Delta t}U^{-1}\right] \left[V^{\dagger}\rho(t)\right] \, .
\end{equation}

\section{A measure of cooperativity}\label{sec:cooperativity}

To quantify the cooperativity of the ensemble of emitters coupled to the same cavity mode, we compare the two situations that are schematically pictured in Fig.~\ref{CoopsFig}. We pump the $N$ quantum emitters at the same rate, $P$. The emitters are either coupled to the same cavity mode ($(a)$), or each of them is coupled to its own resonator ($b$). The output in the photonic channel, which is proportional to the emitted radiation in each case, is compared for the two situations above. The first situation gives rise to an output field (in units of $k$), that is $n(N,\{\delta_i\},\gamma)=\langle a^{\dagger}a\rangle$.  In the second case, we measure the sum of the outputs from each cavity, where each one contains a single emitter, which is written as $n(1,\delta_i,\gamma)=\langle a^{\dagger}_ia_i\rangle$. For a given set of initial conditions, such as pump/dissipation rates, atom-cavity couplings, etc., the system behavior is said to be ``cooperative'' when the two measurements differ, the difference between them giving direct access to the field that is created or annihilated by cooperativity. Then, a cooperativity parameter, or cooperative fraction, can be defined as 
\begin{equation}
C_f=\frac{n(N,\{\delta_i\},\gamma)-\sum_i n(1,\delta_i,\gamma)}{n(N,\{\delta_i\},\gamma)} \, . \label{Cf}
\end{equation}
By construction, this parameter is positive when cooperativity is constructive, while it necessarily assumes negative values for destructive cooperativity. 
It is worth stressing here that the absolute value of $C_f$ as defined in the last equation is not bounded, i.e. it could be arbitrarily large if $\sum_i n(1,\delta_i,\gamma)$ is arbitrarily large. However, this will never be the case in situations one is usually concerned with, in particular the ones treated in the present work. 
The cooperative fraction tends to the limiting value of 1 for maximum constructive cooperativity, and it is ultimately bounded at negative values by the sum of the single emitters output for maximum destructive cooperativity. 
  
 \begin{figure}[t]
\includegraphics[width=\linewidth]{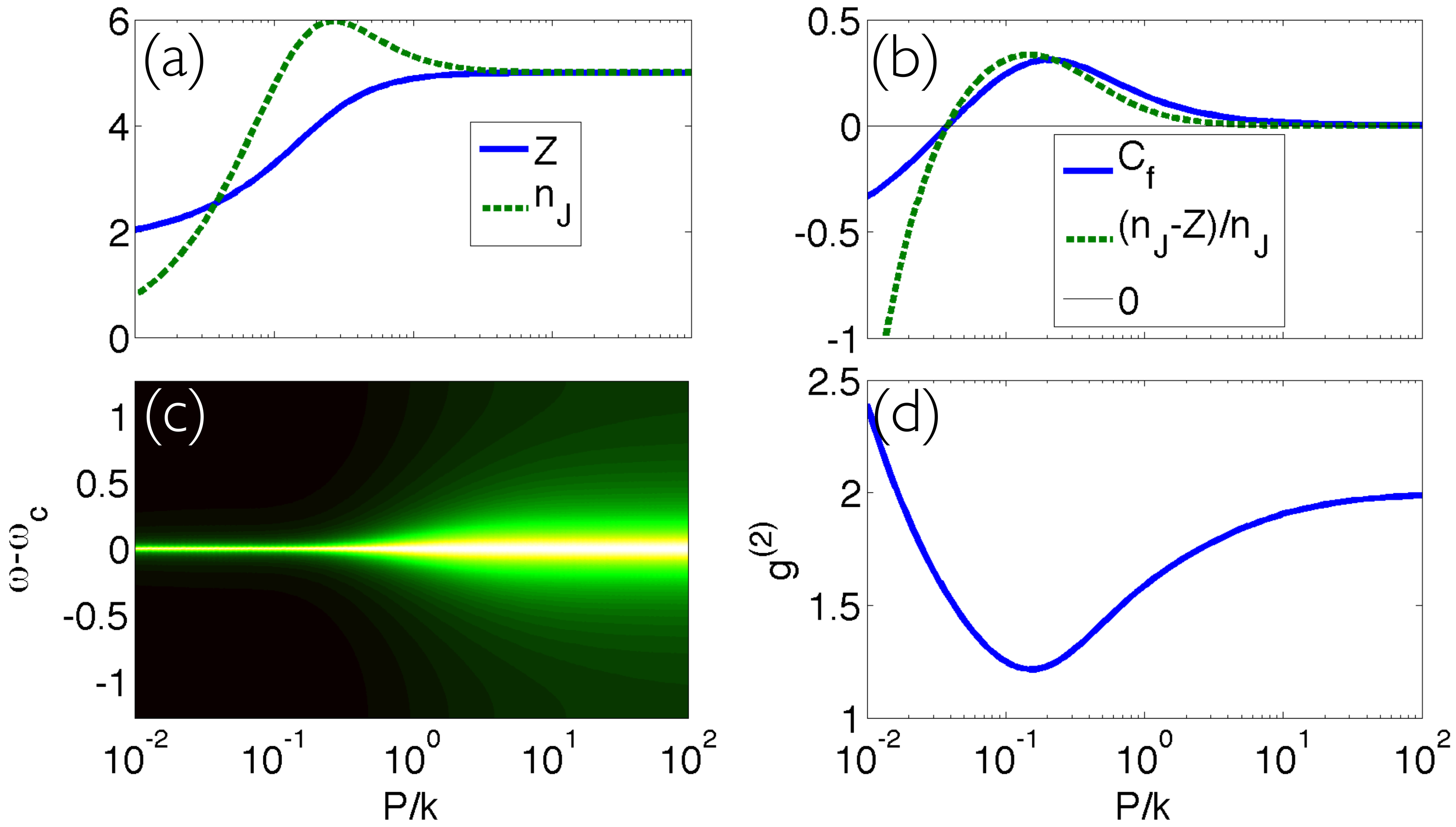}
\caption{Subradiance and superradiance for 5 emitters in the bad cavity regime, $g=0.1k$. 
(a)  Total atomic inversion, $Z$, and average value of the total atomic dipole, $\langle J^\dagger J\rangle$; 
(b) Cooperative fraction compared to $(n_J-Z)/n_J$; 
(c ) Cavity spectrum; 
(d) Second-order coherence function $g^2(0)$. All the data are plotted as a function of the pump rate.}\label{badcav}
\end{figure}  
    
As a first step, we apply our definition of cooperativity to the bad cavity regime of the Tavis-Cummings model, where subradiant and superradiant regimes have already been theoretically characterized \cite{Holland,Auffeves2011}. We set the number of emitters to $5$ and the coupling $g=0.1k$, under which conditions the adiabatic elimination of the cavity mode is appropriate \cite{woggonPRL}: in this regime, $a$ is proportional to the global atomic mode, defined by the collective dipole operator $J=\sum_i \sigma_i$. Cooperativity is usually evidenced by comparing the total atomic population inversion, $Z=\sum_i\langle \sigma_i^{\dagger}\sigma_i\rangle$, to the total atomic dipole $n_J=\langle J^{\dagger}J\rangle$ \cite{woggonPRL}: indeed, it directly compares the field emitted by the atomic ensemble to the field that would be emitted by each two-level system in independent reservoirs. Figure~\ref{badcav}a shows the evolution of these quantities as a function of pumping rate. As already discussed in Refs.~\cite{Holland,Auffeves2011}, a subradiant behavior is manifested at low pumping $P<\Gamma$, where $\Gamma=4g^2/k$ is the relaxation rate of a single emitter. It is due to the efficient optical pumping of the atomic ensemble into its dark states, and gives rise to the emission of highly bunched light \cite{Woggon,Auffeves2011}.  On the other hand, a superradiant regime is reached when $\Gamma<P<N\Gamma$, because of the preferential population of symmetrical Dicke states showing enhanced coupling to the electromagnetic field. Finally, when $P>N\Gamma$, the atomic population is totally inverted, such that each emitter behaves independently from the others \cite{Auffeves2011}. 
Thus, superradiance (subradiance) is characterized by $n_J>Z$ ($n_J<Z$). In Fig. \ref{badcav}b we plot the cooperative fraction, which is negative in the subradiant regime, positive in the superradiant one and goes back to $0$ at high pumping rates, as expected. In the same figure, we also plot the quantity $(n_J-Z)/n_J$, which provides an intuitive measure of cooperativity in the bad cavity regime, based on the analyses already performed in the literature and described above.  As it can be seen, the parameter defined in the present work agrees qualitatively well with former studies, and can be used as a fair marker to describe the transition between subradiant to superradiant regimes, respectively.

As recently demonstrated, a huge ensemble of atoms coupled to a bad cavity can produce extremely coherent light, a phenomenon called "steady-state super-radiant laser" \cite{Bohnet}. In this situation the coherence is to be attributed to the phase locking of the atomic dipoles, and not to the stimulated emission of one given mode of a high quality factor resonator. This result clearly shows the symmetry between steady-state superradiance and stimulated emission, both physical processes inducing an enhancement of the absorption properties of a medium because of bosonic amplification. Nevertheless, in the bad cavity regime, a lasing-like behavior can only take place if the number of emitters is large enough, such that the matter field can be highly excited. This is not the case in the present few emitters situation. This appears in Figs.~\ref{badcav}c and d where the spectrum and second-order coherence function of the cavity field have been plotted. As it can be seen, the second-order coherence does not lock to $1$, and increasing the pump only broadens the spectrum. Hence in the case under study, the lasing character can only come from stimulated emission, which requires a high quality factor resonator. This is investigated in the following Section.

\section{Superradiance-lasing crossover}\label{sec:crossover}

\begin{figure}[t]
\includegraphics[width=0.9\linewidth]{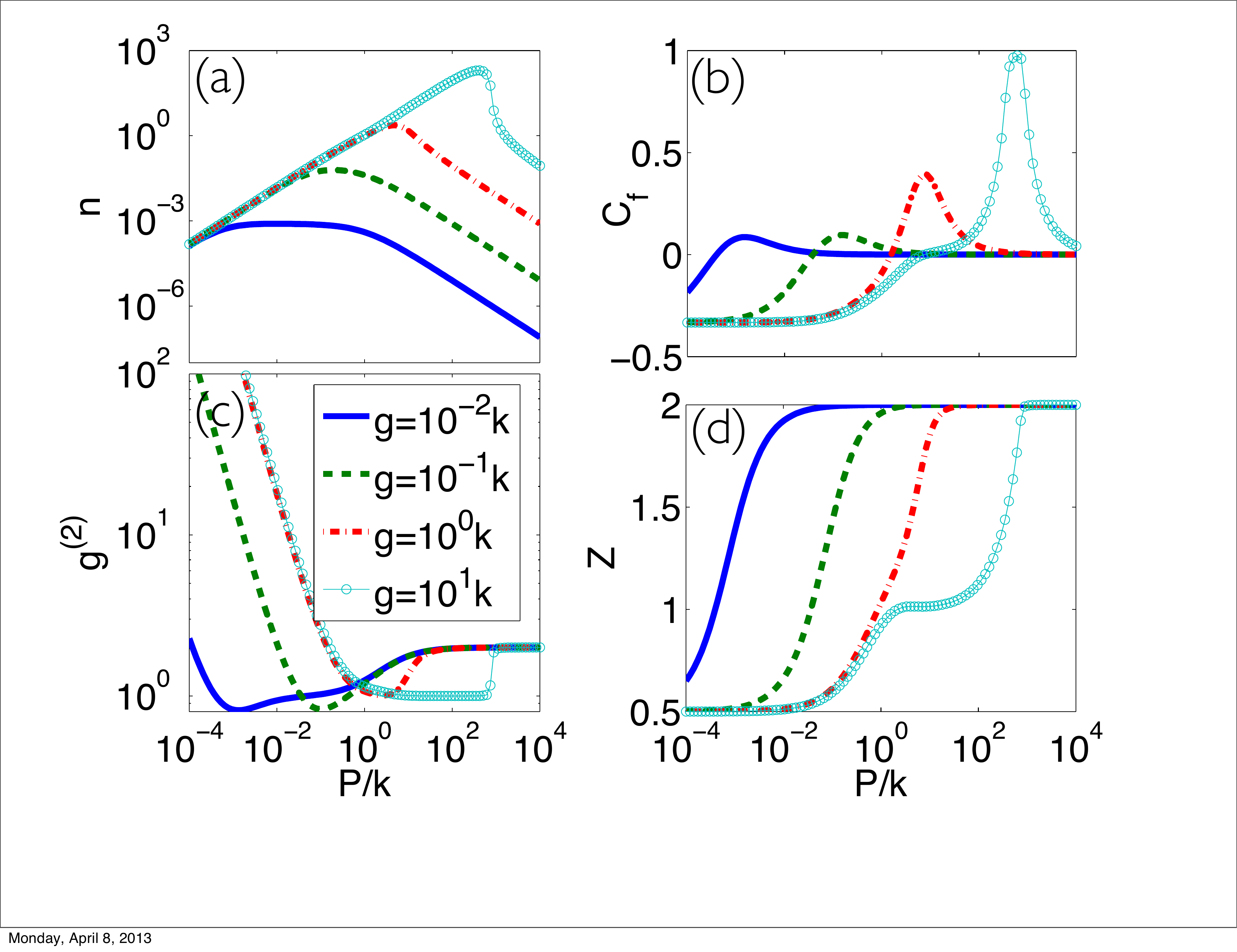}
\caption{The superradiance-lasing crossover is shown for 2 emitters coupled to the same cavity mode: (a) cavity population, (b) cooperative fraction, (c ) second-order coherence function, and (d) atomic inversion. The data are plotted as a function of pumping rate for different values of the emitter-cavity coupling constant. }\label{crossover}
\end{figure} 

In this Section we consider a regime of parameters in which lasing can take place, focussing our analysis on the respective contributions of stimulated emission and superradiance to positive cooperativity. We start from the case of $N=2$, and we progressively increase the light-matter coupling, such that strong coupling regime is eventually reached for each individual atom-cavity system. We have plotted in Fig.~\ref{crossover}a,c and d the cavity population, the second-order coherence function of the cavity field, and the total atomic inversion as a function of pumping rate, for $g$ ranging from $k/100$ to $10k$. In addition to these quantities, which provide usual signatures of lasing, we have also plotted the newly defined cooperativity fraction (Fig.~\ref{crossover}b).
One can mainly distinguish three regimes for cooperativity, corresponding to the cooperative fraction being respectively negative, positive and null. Negative cooperativity still characterizes a subradiant regime, associated with the emission of a bunched light field. 

In the case of positive cooperativity, the analysis performed in the bad cavity regime is valid as long as $g\leq k$. On the contrary, when $g>k$, two different behaviors emerge, as it clearly appears in the plots of Fig.~\ref{crossover}. As a first step, the cooperative parameter $C_f$ becomes positive and remains quite close to $0$. The steady state cavity population increases drastically and the atomic inversion $Z$ is clamped to $1$. These are usual signatures of the lasing regime, confirming that stimulated emission takes place and that the non-linear regime is reached. Simultaneously, a plateau for $g^2(0) \simeq 1$ develops. Indeed, the emission of a Poissonian field induces the locking of the second-order coherence to 1 \cite{loudon_book}. The crossover from non-lasing to lasing regime is captured by $C_f$. In particular, the switch from negative to positive provides a new way to define the lasing threshold in the few emitters case.

In a second step, as we further increase the pump, the cooperativity increases significantly while the system is deeply in the lasing regime. In particular, at the maximum value of cooperativity we have all the typical lasing signatures. In this situation indeed, the medium consisting of a few emitters is still lasing $n(N,\{\delta_i\},\gamma)>0$, while each single-emitter laser has quenched ($n(1,\delta_i,\gamma)\rightarrow 0$). Quenching takes place when the pumping power starts to overcome the effective light-matter coupling strength \cite{alexia2010}. This strength scaling like the number of emitters (see next Section), the few emitters laser quenches for higher pumping power than the individual ones : this is a clear signature of  cooperativity, which is captured by the parameter $C_f$ defined in this study.  This situation can be described in terms of a strongly \textit{cooperative laser}, pointing towards the coexistence of lasing and superradiant characteristics.

Finally the quenching regime appears at high pumping rates. In fact, for lager values of the pump the quenching of the laser begins, and the cooperativity starts decreasing back to zero. Under such conditions, power broadening spectrally decouples the lasing medium from the cavity mode, leading to a decrease of cavity mode population and the switching to a thermal statistics for the emitted field, i.e. $g^2(0) \to 2$. With respect to the cooperativity measure, quenching is clearly manifested by the simultaneous condition $C_f \to 0$.

\begin{figure}[t]
\includegraphics[width=0.9\linewidth]{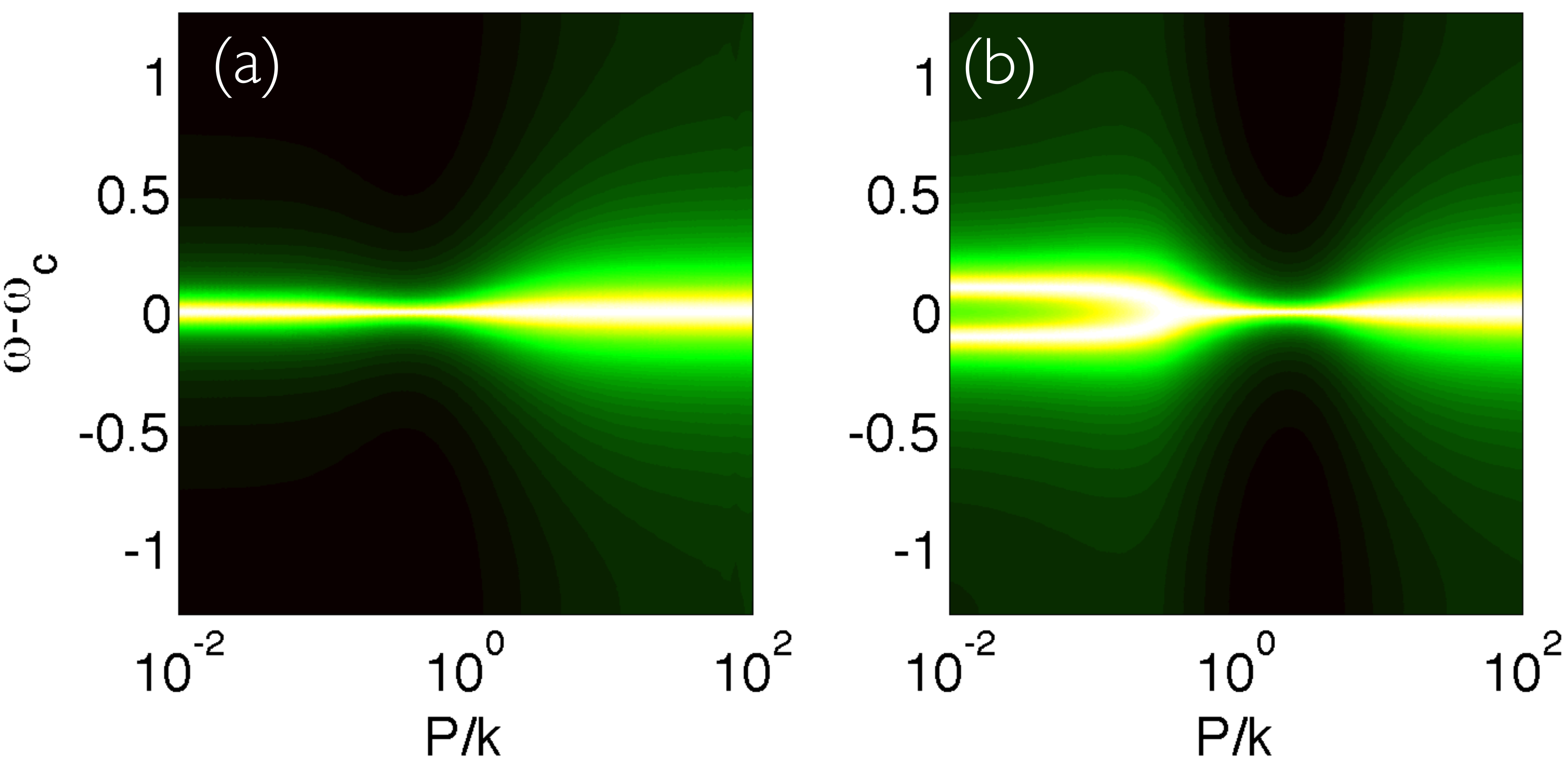}
\caption{The spectral signature of superradiance-lasing crossover for 4 emitters coupled to the same cavity mode: spectrum of the emitted radiation vs. pumping rate for (a) $g=0.2k$, and (b) $g=0.5k$, respectively.}\label{Spec4atTran}
\end{figure}

The signatures of  the crossover from plain superradiance to lasing should also appear in the spectral properties of the radiation emitted through the cavity mode. To confirm this behavior, the results for the calculated spectra are shown in Fig.~\ref{Spec4atTran} for a larger number of emitters (4). The lasing crossover is clearly manifested by a visible narrowing of the emission spectrum in the lasing region. For $g=0.5 k$, at low pumping rates the spectrum is given by the Jaynes-Cummings doublet of each atom-cavity system, under which conditions the lasing narrowing is strikingly evident.

\section{Influence of experimental parameters}\label{sec:parameters}

In this section we focus on the evolution of the cooperativity fraction with respect to the parameters of the ensemble of emitters. In particular, we investigate the influence of the number of emitters, and the robustness of the cooperativity with respect to dephasing and inhomogeneous broadening.

\subsection*{Cooperativity and number of emitters}

\begin{figure}[t]
\includegraphics[width=\linewidth]{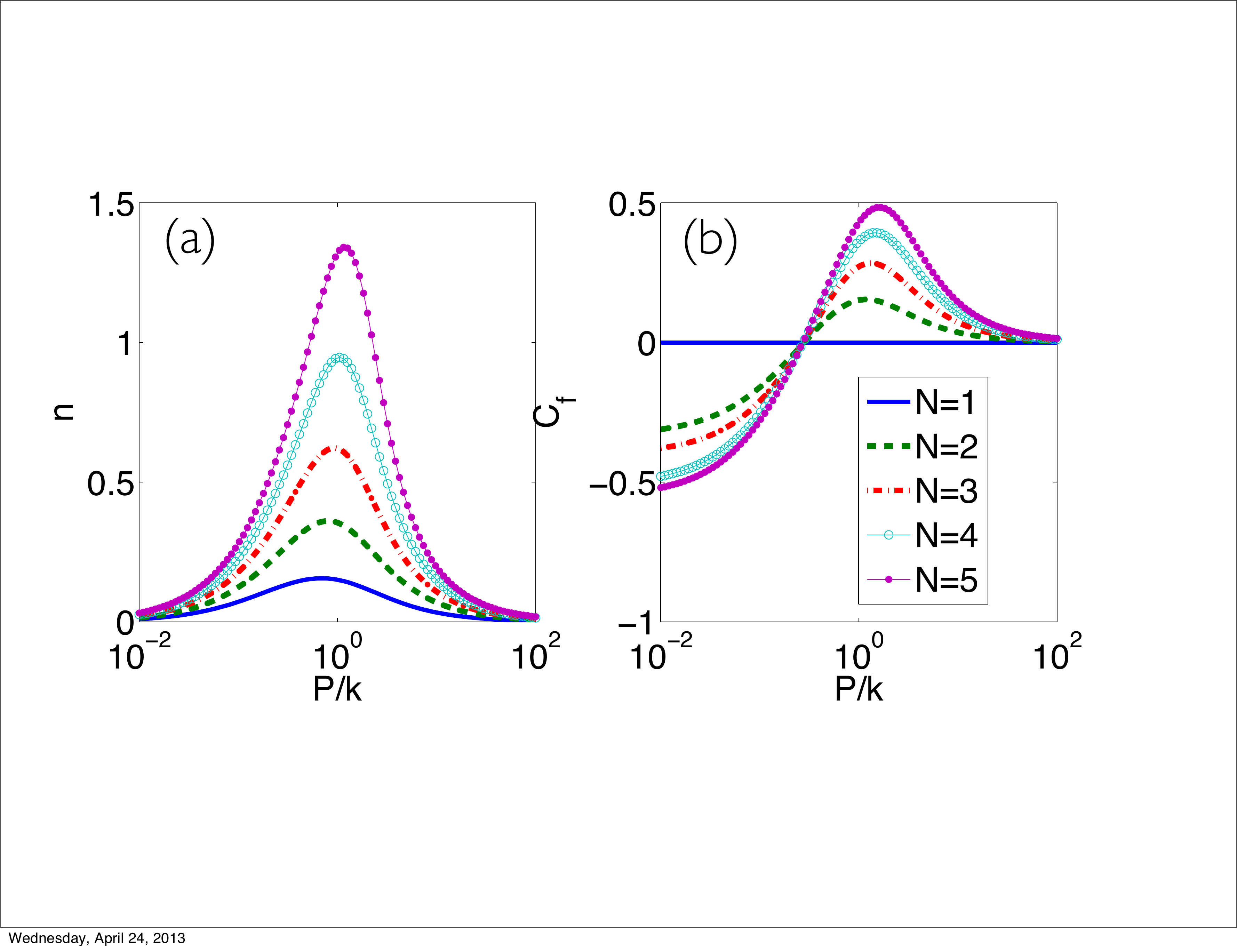}
\caption{ Influence of the number of emitters in the bad cavity regime, $g=0.3k$: (a) cavity population, and (b) cooperativity fraction.}\label{SupperN}
\end{figure}

The analysis of the emission properties of a mesoscopic number of emitters in a cavity by increasing the number one by one is a fruitful bottom-up approach, which was already taken in \cite{Auffeves2011} to investigate the statistical properties of the emitted radiation, and paves the way to a novel description of the quantum-classical boundary. The results for the system parameters in the bad cavity regime are shown in Fig.~\ref{SupperN}. The maximum of the cavity population plotted in Fig.~\ref{SupperN}a increases drastically with the number of emitters, a behavior already emphasized in \cite{Holland}, where such saturation value has been shown to evolve as $N^2$. In Fig.~\ref{SupperN}b, we can appreciate the corresponding evolution of the cooperativity fraction. In the subradiant regime, $C_f$ decreases with the number of emitters: this can be interpreted by noting that the low excitation Dicke states are mainly populated on average, for which the decay rate typically scales as $N$. As a consequence, the larger $N$, the faster the relaxation, which leads to a lower excitation of the matter field at equal pumping rate and fully justifies the behavior of $C_f$. As expected, $C_f$ switches from negative to positive values for $P=\Gamma$, which does not depend on the number of emitters. On the contrary, and as it also appears in the Figure, the system returns to an independent-like behavior when $P=N\Gamma$, which occurs at larger and larger pumping rates on increasing $N$.

\begin{figure}[t]
\includegraphics[width=0.8\linewidth]{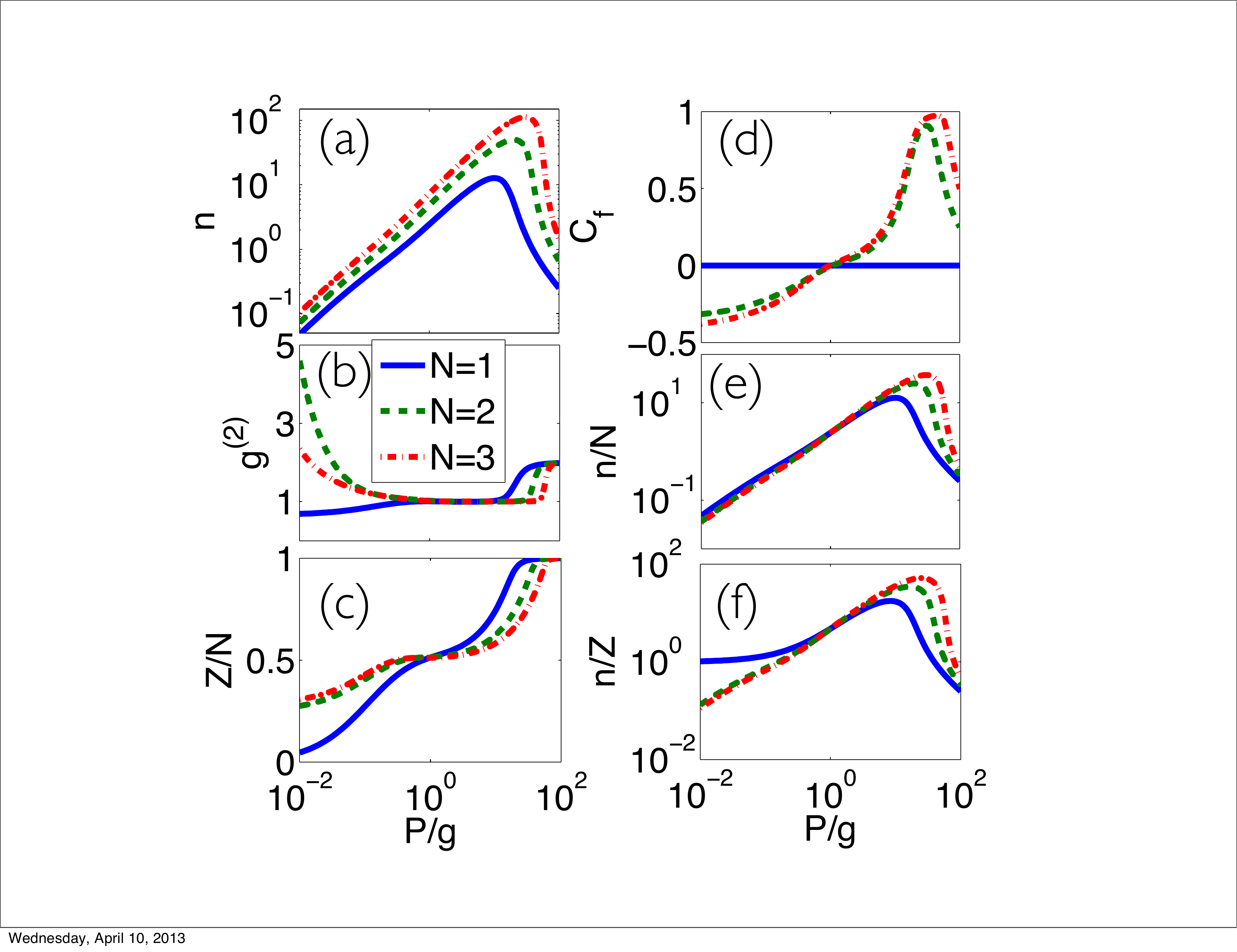}
\caption{ Subradiance and lasing in the strong coupling regime for increasing number of emitters, i.e. $g=5 k$: 
(a) cavity population, 
(b) second-order correlation function,  
(c) population inversion per emitter, 
(d) cooperative fraction,
(e) cavity population per emitter, 
(f) cavity population per excited emitter. The data are shown for one emitter (blue continuous), two (green dashed) and three  (red dashed-dotted) emitters, respectively. All the data are plotted as a function of the pumping rate.}\label{123Lasing}
\end{figure}

In the same spirit as above, we revisit now these properties in the case where the resonator has a good quality factor, so that each emitter is individually strongly coupled to the cavity mode ($g=5 k$). We have investigated the signatures of lasing for a medium consisting of a few identical quantum emitters, ranging from $1$ to $3$. Usual quantities, namely cavity population $n$, second-order correlation at zero time delay, $g^2(0)$,  and population inversion (normalized to the number of emitters, $Z/N$) are plotted on the left side of Fig.~\ref{123Lasing}.  As previously, the regimes of subradiance, superradiance/lasing and quenching can be clearly identified. 
Subradiance is manifested in the low pumping regime, by a negative cooperativity fraction and oscillations of the second order coherence function between even and odd numbers of emitters. These oscillations have been thoroughly analyzed and already discussed in Ref.~\onlinecite{Auffeves2011}, essentially related to differences in the Hilbert space geometry of Dicke states participating in the driven/dissipative quantum dynamics, and depend on the parity of $N$. In particular, these oscillations are the signature of the energy spreading of the Dicke eigenstates, which depends on the parity of the number of atoms~\cite{Auffeves2011}. Another signature is the excitation of the matter field even in the low pumping case, as already mentioned: such an effect reflects the optical pumping in the dark states~\cite{Holland} and clearly appears in the plot of the population inversion. It can also be seen that the extension of the lasing regime as a function of the pumping strength is larger the larger is $N$, as saturation and quenching take place at larger pumping rates. For completeness, we have also reported the results for the single-emitter case, which has been thoroughly investigated in the literature~\cite{Oneatomlasers,QuantumopticalmasterequationsTheoneatomlaser,Spectralpropertiesoftheoneatomlaser,delValle2009,alexia2010,delValle2011}. Again, both effects can be accounted for by the increase of the effective light-matter coupling, which scales as $\sqrt{N}$.
On the right hand side of Fig.~\ref{123Lasing}, we plot the cooperative fraction, the cavity population per emitter, $n/N$, and the cavity population per excited emitter, which is defined as $n/Z$. 
The influence of $N$ on $C_f$ is quite straightforward to analyze: increasing $N$ increases the magnitude of the cooperativity fraction. In the subradiant regime, $C_f$ is more negative, as in the bad cavity case. In the superradiant regime, the maximum $C_f$ occurs at larger pumping, and it reaches larger values: this corresponds to the saturation and quenching of $N$ cooperative emitters occurring at larger pump power as compared to the single-atom laser. The maximum cooperativity is reached in this very region where the cooperative medium is still lasing, while the single emitter medium has quenched. Switching from negative to positive cooperativity does not depend on $N$, as in the low Q resonator case. The same evolution can be seen in the parameter $n/N$. 

Finally, we have studied the quantity $n/Z$ (see Fig.~\ref{123Lasing}f), which is the typical absorption per emitter. Its meaning in the single atom case is clear:  in the spontaneous emission regime, the atomic inversion and the cavity population scale like the pump power, hence the ratio $n/Z$ is constant and locked at the spontaneous emission rate. When the non-linear regime is reached, the population clamps while the cavity population is still increasing. Hence the parameter $n/Z$ increases linearly with the pump power because of stimulated emission, a feature that clearly appears in the plot. Such behavior would also appear for a standard lasing medium made of distinguishable atoms. On the contrary, for the microlaser investigated here, the switching from linear to non-linear behavior is blurred out as soon as the medium involves more than a single emitter. As it can be seen in the plot, the average absorption per emitter continuously increases with respect to the pump power. Indeed, here the increase in the absorption is the result of two contributions: before the lasing threshold, it is due to the displacement of the equilibrium in the Dicke states phase space toward higher excitation states characterized by higher coupling to the light field (steady state superradiance). When lasing takes place,  the absorption increase comes from stimulated emission. This behavior is a major difference between standard lasing media made of distinguishable atoms and the specific medium investigated here.  
At the highest powers the drop-off in $n/Z$ is attributed to quenching, which describes the decrease in the absorption of the atomic medium when the pump power becomes too high. This effect is due to the broadening of the atomic emission line at high excitation power \cite{alexia2010}.

\subsection*{Cooperativy vs. dephasing/inhomogeneous broadening}

\begin{figure}[t]
\includegraphics[width=\linewidth]{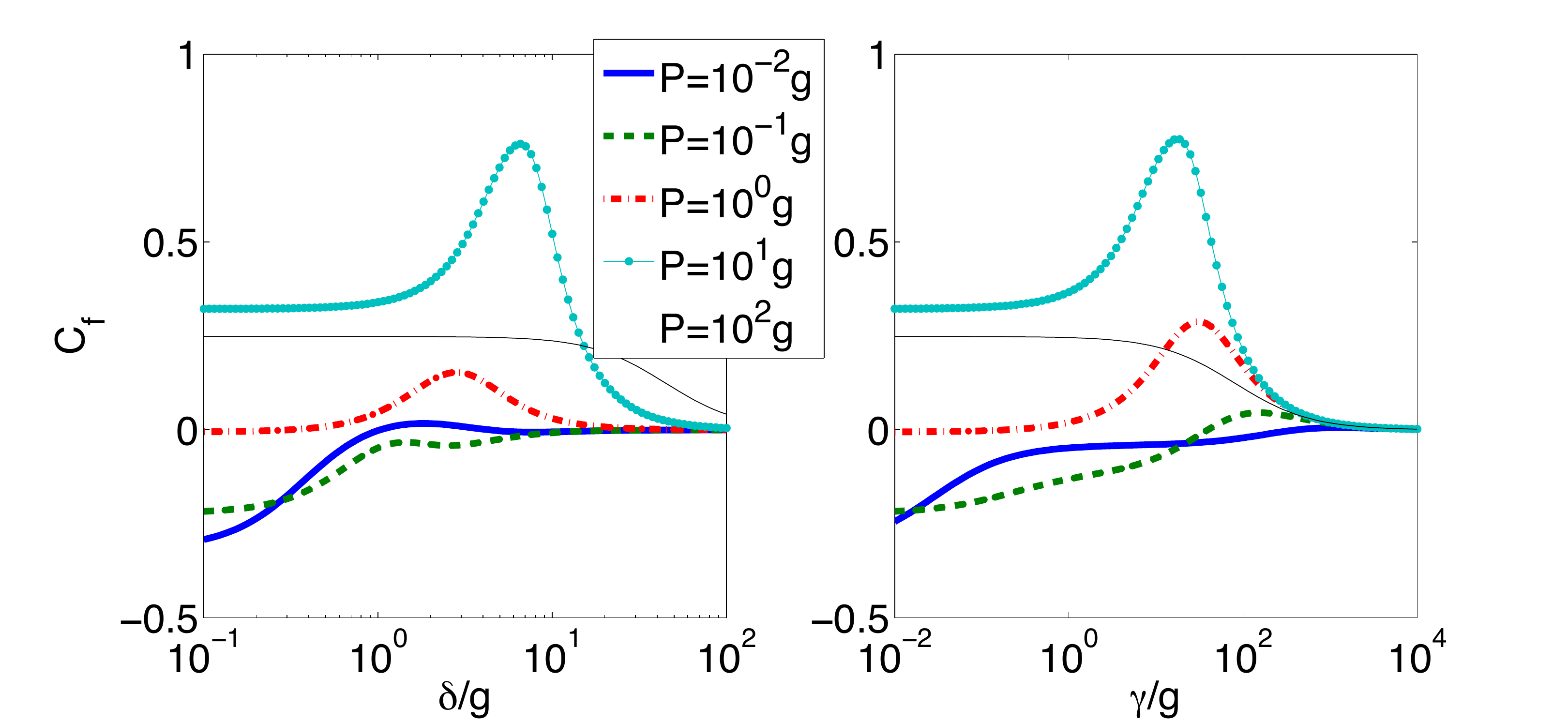}
\caption{Cooperativity fraction in the two-emitter case: (a) as a function of (symmetric) detuning from the cavity resonance frequency, (b) as a function of pure dephasing rate (and zero detuning). We considered the strong coupling regime $g=5 k$.}\label{CoopFrac2atomsDetuEQ}
\end{figure}

So far we have assumed the quantum emitters to be identical, on resonance with the cavity mode, and non-dephased. Now we relax this latter constraint and analyze how detuning (i.e. inhomogeneous broadening) and dephasing affect the cooperativity. Both of these effects are relevant in solid state implementations of the model studied in this work, such as in a quantum dot laser \cite{strauf06}. We consider a system of two emitters equally detuned from the single-mode cavity frequency, and compute their cooperativity as a function of detuning for different regimes of the pumping rate. The results are plotted in Fig.~\ref{CoopFrac2atomsDetuEQ}a. 
As it appears in the plot, increasing the detuning may actually increase the cooperativity,  even though the total atom-cavity coupling decreases. This increase is due to the fact that the cavity population in the single emitter is more sensitive to detuning. Thus, while the single emitters are being decoupled from their cavities the cooperative emitters sustain the cavity population. For low pumping regimes the system may be driven from subradiant to superradiant regime. We refer to this positive cooperative regime as superradiant instead of lasing, since it lacks other characteristic features of lasing, such as spectral narrowing (not shown here). For larger values of the pumping rate, the system behaves as a regular laser, as already established. In this regime the cooperativity also presents a maximum as a function of the detuning even though, once again, the effective atom-cavity coupling only decreases. Finally and as expected, for large detuning the cooperativity tends to zero.
 
Interestingly, a similar behavior is found by adding pure dephasing to the emitters, as shown in Fig.~\ref{CoopFrac2atomsDetuEQ}b. However, it is clear that the cooperativity is much more sensitive to detuning rather than dephasing. 
We also point out that detuning the emitters asymmetrically (not shown here) with respect to the cavity frequency, or non equally detuning them, usually leads to less cooperativity as compared to the case in which the emitters are symmetrically detuned around the cavity mode frequency. This sensitivity is due to the fact that the resonator couples to the symmetric "brilliant" state of the two emitters, which remains identical if the two emitters are detuned symmetrically with respect to the mode.

 \section{Conclusion}
 
In conclusion, we have introduced a new operational quantity to measure the degree of cooperativity in the emission characteristics of a system of few incoherently pumped emitters coupled to a single-mode cavity. It is defined by comparing the photon emission of an ensemble of emitters coupled to the same cavity mode with the overall emission of the same emitters each one individually coupled to its own resonator mode. We have shown how such a quantity is able to quantitatively describe the crossover between steady state subradiance, superradiance, and lasing. We have analyzed the effects of inhomogeneous broadening and pure dephasing on cooperativity, which might be relevant in solid state implementations of this model. We have shown that in the good cavity regime, a few indistinguishable two-level systems provide a new type of lasing medium as compared to an ensemble of distinguishable emitters. These results are quite promising for emerging experiments in mesoscopic quantum optics, where a bottom up approach allows the possibility to sequentially add an increasing number of emitters coupled to the same cavity mode.

\acknowledgements
The authors acknowledge Brazilian Agencies CNPq and Fapemig, the NanoSci EraNet grant "LECSIN," the Fondation Nanosciences of Grenoble and ANR-PNano "CAFE", and the Italian Ministry of University and Research through Fondo Investimenti Ricerca di Base (FIRB) "Futuro in Ricerca" project RBFR12RPD1 for financial support.

\end{document}